# You Can Beat the "Market": Estimating the Return on Investment for National Hockey League (NHL) Team Scouting using a Draft Value Pick Chart for the NHL

Michael Schuckers[1*], Steve Argeris[2]

St. Lawrence University[1], Statistical Sports Consulting, LLC[1], Hunton & Williams LLP[2], George Mason University[2], and Georgetown University[2], michael.schuckers@statsportsconsulting.com

# You Can Beat the "Market": Estimating the Return on Investment for National Hockey League (NHL) Team Scouting using a Draft Value Pick Chart for the NHL

**Abstract:** Scouting is a major part of talent acquisition for any professional sports team. In the National Hockey League (NHL), the 'market' for scouting is set by the NHL's Central Scouting Service which develops a ranking of draft eligible players. In addition to the Central Scouting rankings, NHL teams use their own internal scouting to augment their knowledge of eligible players and develop their own rankings. Using a novel statistical approach we show in this paper that the additional information possessed by teams provides better rankings than those of Central Scouting. Using data from the 1998 to 2002 NHL drafts, we estimate that the average yearly gain per team from their internal scouting is between $1.8MM and $5.2MM. These values are consistent across the three measures of player productivity that we consider: cumulative Games Played, cumulative Time On Ice and cumulative Goals Versus Threshold where we have aggregated these metrics across the first seven years post draft. We used this time frame since teams generally retain rights to their draft picks for seven years. Further, we find that no individual team outperformed the others in terms of draft performance. One byproduct of our analysis is the development of a Draft Value Pick Chart to assess the worth of an individual selection.

**Introduction and Background**

The Edmonton Oilers selected Jordan Eberle, a center, with the 22$^{nd}$ pick in the 2008 National Hockey League (NHL) Entry Draft. Three picks later, at 25, the Calgary Flames selected another center, Greg Nemisz. This was not the obvious order of things at the time. Going into the draft, the NHL's Central Scouting Service (CSS) ranked Eberle as the 33$^{rd}$ best North American skater while Nemisz ranked 22$^{nd}$ in that same category; in other words, the league's own amateur scouts ranked Eberle as a second-round pick at best, even if you ignored the dozens of North American goaltenders and Europeans available. Since the draft, Eberle has played 275 NHL games, while Nemisz has played in 15. The Oilers presumably had additional information from their internal scouting staff about why to pick Eberle over Nemisz, as well as other centers that were available with the 22$^{nd}$ pick, including Daultan Leveille, who has yet to play in the NHL, and Derek Stepan, who has played nearly 300 games and represented the United States at the 2014 Olympics.

Sometimes the information teams possess misleads them. Infamously, the New York Rangers used the 12$^{th}$ overall pick in 2003 to select Hugh Jessiman, a right wing from Dartmouth ranked the 20$^{th}$-best North American skater by CSS. The Rangers preferred Jessiman to a virtual all-star team's worth of talent (Dustin Brown, ranked the No. 2 North American skater by CSS, along with Brent Seabrook, Zach Parise, Ryan Getzlaf, Brent Burns, Ryan Kesler, Mike Richards and Corey Perry). Jessiman has since played two NHL games, and signed a contract for the 2014-15 season with the Capitals, albeit in Vienna, not Washington.

CSS was formed by former NHL general manager Jack Button in 1975 as a service to NHL clubs to scout amateur prospects and later administer the NHL's annual combine, which invites the top 100 prospects to come to a centralized location for physical examinations, light drills, and interviews with teams. (Malloy, 2011). Other leagues have had some form of centralized scouting as a cost-saving service to their member clubs, such as the Major League

Baseball Scouting Bureau, begun in 1974, but no leagues have placed such a public emphasis on its scouting service's rankings as the NHL, which publishes various CSS lists several times per year to great fanfare. The CSS employs nearly 30 scouts in the field, more than any team's in-house department, including six European-based scouts via Goran Stubb and his Finland-based European Scouting Services. (Morreale, 2011; Shoalts, 2010). Each NHL team has available to them information about possible draftees from the CSS as well as information compiled by their own staffs. The outcomes of draft decisions like those of Edmonton and Calgary are full of variation—more than half of all players selected in the NHL Entry Draft never play a single game in the NHL.

In this paper, we will look at the quality of rankings by the CSS and draft order by team in order to evaluate and quantify the value of the additional information that teams have. In doing so, we will estimate the average annual return that teams get from their internal scouting. This is analogous to an approach used in evaluating the success of portfolio managers in finance—the search for "alpha", or risk-adjusted returns in excess of those readily available in a market index. (Jensen, 1967). Were the "index", in this case the freely available CSS rankings, comparably successful at picking talent after adjusting for the cost of teams running their own scouting departments, it would endorse a more passive approach to talent evaluation, similar to that of John Bogle when he launched the first index mutual fund at Vanguard in 1975. If nothing else, the CSS can serve teams as "a naïve model against which their in-house gunslingers can measure their prowess" as Paul Samuelson (1974) wrote of a market index fund and money managers.

The data that we analyzed here is from the five NHL Entry Drafts from 1998 to 2002. For each player selected we have their selection number, their position, time on ice (TOI), the games played (GP) in the NHL, the ranking by player type by CSS and their career goals versus threshold (GVT). Teams have the rights to players for at least the first seven years after they are drafted.[1] Consequently, we will focus our analysis on player performance during this period after a player is drafted. Below we will use the total of each metric for the first seven seasons after they were drafted. Since for all of the players in our sample this includes the 2004-5 lockout season, we will use eight total years to capture seven seasons worth of data for each player. These data were compiled from nhl.com, hockey-reference.com, and eliteprospects.com. We used the CSS final rankings released prior to the draft. These rankings were obtained via contemporaneous media accounts accessed via Internet searches and the Lexis/Nexis database. A player's position was recorded as either center (C), defensemen (D), forward (F), goalie (G), left wing (L) or right wing (R). In the analyses that follow we will categorize players as forwards (F) which includes C's, L's, and R's, defensemen (D) or goalies (G). The selection number for a player is the place in the draft order that they were selected. That is, the 10th player selected will have a selection number of 10, the 100th player selected will have a selection number of 100, etc.

---

[1] NHL free agency is slightly more complicated, but generally speaking players reach free agency at age 27 or after seven NHL seasons, whichever comes first. The age limit was gradually reduced after the 2004-5 lockout from 31 to 27 for the players in our sample, but they did have collectively bargained options to have some access to market forces while under their drafting team's control such as salary arbitration and restricted free agency, the latter of which allows for players to sign with another team in exchange for that team's picks in the following NHL draft.

CSS ranks players by category, either North American or European, and by whether they are a skater or a goalie. Thus, CSS produces four separate rankings without a correspondence between them for comparison purposes. To better utilize the CSS rankings, we employed Iain Fyffe's "Central Scouting Integratinator" (CESCIN) metric, Fyffe (2011). CESCIN takes the rankings of players by CSS within their given category and multiplies them by a factor based upon historical draft selection records. For players who were drafted but not ranked by CSS, we gave those players values of CESCIN that was larger than the values produced by the original CESCIN. Next, we took the values generated by CESCIN and ranked them to produce the CSS orderings we use in the rest of the paper. Three response metrics are considered below: time on ice (TOI), games played (GP) and goals versus threshold (GVT). GVT is a metric of player value created by Tom Awad that allocates value in team performance among the individuals on a given team, Awad (2009). The units for GVT are goals so that a GVT of 7.5 credits a player with producing 7.5 goals over replacement level. Since players can have negative GVT, we give players who never played in the NHL a value of GVT below the lowest value in our database. Our focus is on the top 210 selections since that is the length, over seven rounds, of the six most recent NHL Entry Drafts (2008 to 2013). Thus, for each year we have 210 observations except for 2002[2].

For the data that we are considering, there were 595 were forwards, 332 as defensemen (D's), and 122 as goalies (G's). For GP, 54% of the players selected never played a game in the NHL. Table 1 has statistical summaries of TOI, GP and GVT. For GVT, there were no values of that metric for players that did not appear in an NHL game. Among those who did play at least one game the worst GVT that was calculated was -27.7. For completeness and comparability, we chose to give those who did not appear in an NHL game a GVT of -30. We selected this value to be below the other values in our data and to permit a more complete analysis. For goalies, we gave them 20 minutes of TOI for every game in which they appeared. Again this was done to facilitate comparison across positions. We considered alternatives to the 20 minutes for goalies as well as weighting minutes for defensemen differently than forwards; however, the results presented here were not materially changed by these alternatives. To gauge team preferences for players we will use the selection at which the player was taken. For assessing how CSS ranked players we will use CESCIN values.

Table 1: Statistical Summaries of First Seven TOI, GP and GVT

|  | Median | Mean | 75th percentile | Max | Std. Deviation |
|---|---|---|---|---|---|
| TOI | 0 | 1037 | 684 | 13880 | 2053 |
| GP | 0 | 69 | 81 | 553 | 124 |
| GVT | -30 | -12 | 0.3 | 114 | 24 |

---

[2] In 2002, the 123rd pick in the draft was invalidated when the Edmonton Oilers selected a player who was ineligible to be drafted.

There has been a good deal of previous work on evaluation of NHL draft picks. Much of this work has focused on the value of an individual selection. Johnson (2006), Tango (2007), Awad (2009), Gregor (2011), Schuckers (2011) and Tulsky (2013) have all looked at methods for evaluating what an individual draft pick is worth in terms of a measure of value. Many of the critiques of team drafting are that for a given selection a player whose future performance exceeded the current selection was often available. See, for example, Tingling (2011). Given the difficulty with projecting the future performance, it is important to focus on the trends rather than individuals. Clearly there is monotonicity in average performance of players versus draft selection and clearly there are long tails to the distributions of player performance for a given selection. The focus of our analyses is the value that teams get from their scouting departments. It is difficult to quantify the value of CSS since teams have information both from CSS and from their own internal scouting departments. Serge Savard, a Hockey Hall of Fame player who won two Stanley Cups as a general manager for the Montreal Canadiens, acknowledged that the CSS rankings heavily influenced their own, and the difficulties of comparing talent across multiple leagues, even within Canada:

> We were wrong on the first round maybe 50 percent of the time. That's mainly because of Central Scouting. When Central Scouting comes out with their first-round list, all the scouts think, "Oh, Christ, I better get this player in my list or I'll look bad." [All the scouts'] lists are similar because of Central Scouting. I only had one guy, Rick Taylor, who didn't care about Central Scouting's list and his list was so different than the others.... How come we missed Luc Robitaille? One of my scouts, Rick Taylor, had Luc Robitaille [rated to be drafted] in the first round and nobody else had him in the top five rounds. The other scouts down-played Taylor. They said, "You only see Quebec. You don't see Ontario. You don't see the West. You don't see college. You don't see Europe." So scouting is a tough thing to do. (Farris, 2011).

We propose that the CSS represents a suitable, if crude, proxy for a benchmark index such as the Standard and Poor's 500. Teams could, theoretically, get rid of their amateur scouting department and rely on CSS's rankings, as it typically ranks more than the 211 players drafted.[3] Below we will look at the difference in how teams rated players and how CSS rated players to get an idea about the value added by team scouting staffs.

The number of scouts employed by teams varies considerably. For example, prior to the 2013-14 season the New York Islanders had 11 individuals with scouting responsibilities while the Toronto Maple Leafs listed 23 on their respective webpages. One small market U.S.-based team estimates that they spend approximately $2 million on their annual scouting budget. This is consistent with the Phoenix Coyotes' 2009 income statement disclosed as part of its bankruptcy, which included a line item of $1.4 million for "scouting operations", presumably encompassing both amateur and professional scouting, (*In re Coyotes Hockey LLC*, 2009). The Phoenix Coyotes' media guide for the following season (2009-10) listed a 10-member scouting department, including management. Teams have attempted heavy cost-cutting measures, such as the Buffalo Sabres' 2006 overhaul of its well-regarded scouting department to heavily emphasize video scouting, (Joyce 2008). Other teams, both smaller- and larger-budget, have viewed scouting budgets as a competitive advantage. The New Jersey Devils, traditionally a lower-revenue team, have regularly employed more than 20 scouts, while Brian Burke increased the large-revenue Maple Leafs' scouting budget upon taking over the team in 2009, attempting to "exploit that advantage" that there are no league-mandated constraints on scouting budgets, (Shoalts 2010). Similarly, upon his purchase of the Sabres in 2011, new owner Terry Pegula saw scouting budgets as a point of competitive advantage for larger-budget teams: "There is no salary cap in the National Hockey League on scouting budgets and player-development budgets." (Klein and Hackel, 2011).

---

[3] Though we note that this would be rather inadvisable for strategic reasons, including but not limited to the fact that teams would have little idea who other teams' top prospects were for trade value as the players progressed.

In the rest of the paper we begin by considering how often CSS rankings and team draft order were able to optimally or nearly optimally selected the best player at a given selection. We find that teams outperform the CSS rankings. Next we consider a non-parametric LOESS regression following Schuckers (2011) of our performance metrics onto the player orderings, both CSS and team. This approach also finds that, on average, teams outperform the CSS. This work is the basis for a new Draft Value Pick Chart (found in the Appendix) based upon TOI for a player's first seven seasons. Finally, we consider a novel approach that looks at the relationship in the rank orderings and relative value of our performance metrics. Since this final approach is conditional at the individual level, it is the most relevant and informative. We then calculated the excess value above the CSS rankings that teams get in terms of GP and GVT and these are roughly $4 million dollars per year.

Table 2: Comparison of Performance of CSS and Actual Draft Orderings

| Metric | Ordering | Percent of optimal ordering | Percent of nearly optimal ordering |
|--------|----------|-----------------------------|-------------------------------------|
| TOI    | CSS      | 14%                         | 19%                                 |
|        | Team     | 20%                         | 32%                                 |
| GP     | CSS      | 4%                          | 17%                                 |
|        | Team     | 11%                         | 30%                                 |
| GVT    | CSS      | 4%                          | 10%                                 |
|        | Team     | 10%                         | 14%                                 |

**Quality of Central Scouting Draft Order**

In this section we assess the ability of CSS to correctly order the possible draft selection based upon CSS's own ordering of draftees. To evaluate this we looked at the percent of times that the ordering by CSS as reflected in CESCIN resulted in the optimal ordering at a given position, either C, D, F, G, L or R. Note that this differs from the way the CSS ranks players which combines centers, forwards, defensemen, left wings and right wings into skaters. We also considered the percent of times that the CSS came within approximately one-half standard deviation of the optimal choice at a given position. We determined if a selection was **optimal** or **nearly optimal** by considering all of the remaining draftees in a given year at the same position as the selected player. If the selected player had the highest metric among all other *available* players, then that player was considered the optimal selection. To be nearly optimal the player had to be within half of a standard deviation (SD) of the highest metric for all other *available* players. For GP, the standard deviation (SD) of the players taken in the first 210 selection was 215 games, while the standard deviation of those same players for GVT was 21. Table 2 has the results of this analysis for TOI, GP and GVT. Overall it is clear that team ordering (based upon the actual draft) is better than CSS ordering. Simply choosing the best available player (optimal ordering) happens about 8% of the time, on average, using CSS and about 14% of the time with team ordering. Ordering players so that the current selection is nearly optimal happens about an average of 14% of the time for Central Scouting. For teams this latter value is about 26%. A further analysis of these data indicates that the advantage for team ordering is persistent across rounds. These results can be found in Table 3.

**Comparison of Average Performance by Player Ordering**

Above, we have concentrated on the optimal or nearly optimal decision at a given selection. We next looked at the impact of these selection criteria on the average outcome variables per draft selection. To evaluate this impact we looked at the relationship between our response metrics (TOI, GP and GVT) and player ordering. Figure 2 has plots of these relationships. As before, ordering for CSS is done based upon CESCIN while ordering for teams is from the actual draft selections. To estimate these relationships we use LOESS regression as was done in Schuckers (2011) for National Football League data. LOESS regression is a flexible non-parametric methodology for locally smoothing the response at each value of the predictor (selection).

Table 3: Comparison of Drafting Performance by Ordering and by Round

| ***GP*** | | | | | |
|---|---|---|---|---|---|
| ***Ordering by*** | Team | | | CSS | |
| Rounds | Optimal | Near Optimal | | Optimal | Near Optimal |
| 1 to 3 | 6% | 15% | | 2% | 5% |
| 4 to 7 | 12% | 30% | | 9% | 27% |

| ***GVT*** | | | | | |
|---|---|---|---|---|---|
| ***Ordering by*** | Team | | | CSS | |
| Rounds | Optimal | Near Optimal | | Optimal | Near Optimal |
| 1 to 3 | 6% | 10% | | 3% | 4% |
| 4 to 7 | 12% | 18% | | 10% | 12% |

| ***TOI*** | | | | | |
|---|---|---|---|---|---|
| ***Ordering by*** | Team | | | CSS | |
| Rounds | Optimal | Near Optimal | | Optimal | Near Optimal |
| 1 to 3 | 16% | 21% | | 11% | 14% |
| 4 to 7 | 21% | 39% | | 18% | 37% |

In the leftmost graph of Figure 2 we have plots of the LOESS regressions for predicting TOI based upon the draft selection order from CSS (blue) and from the actual draft (red). Our expected or predicted values from both of these curves are very similar. Both start at roughly the same value for the first selections and decrease steeply until approximately the 40th selection, then less steeply until the 120th selection after which they are roughly flat. The difference between the two curves is that the draft selections outperform the CSS rankings from about 40 to 100. Currently this corresponds to roughly the early second round to the end of the third round. These are the locations where, on average, team selections are better than the ordering from Central Scouting. Central Scouting does better than teams over the last 35 selections. Moving to GP and GVT, we see a similar shape to the smoothed prediction relationship that we found for TOI. Although somewhat hard to judge due to the different scaling of the y-axes, the overall shape of these curves is quite similar.

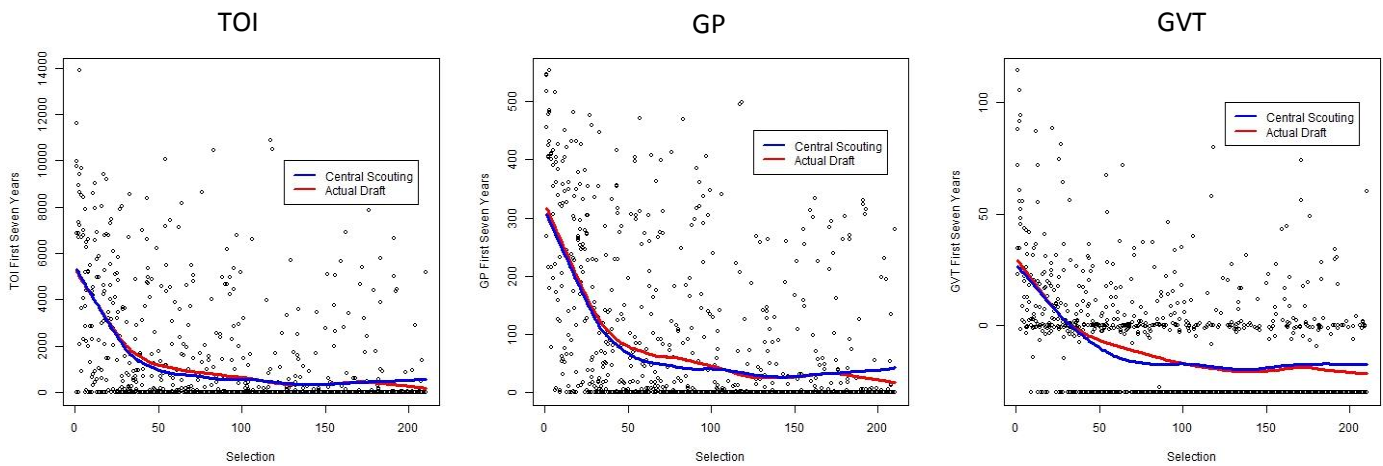

**Figure 1: Average Performance Comparison for CSS and Team Draft Ordering: (l to r) TOI, GP, GVT**

**Rank Differential Comparison**

While the above analysis gives a sense of how CSS and team scouting group perform in their rankings on average, those comparisons and rankings do not condition on some important factors such as the individual player and their position. To account for these we next consider an analysis that looks at the differential between each player's actual selection and their CSS ranking. Below we will refer to this as rank differential (Δ rank). For example, Rico Fata was the 6th overall selection by the Calgary Flames in the 1998 draft. CESCIN has Fata as the 13th ranked player. Consequently our rank differential for Fata would be -7 meaning he was taken seven places ahead of where CSS ranked him. Players with negative rank differential were taken earlier than CSS ranked them and players with positive rank differential were taken later than CSS ranked them. 56% of selections were rank differential positive, 43% were negative and 1% were zero. If team scouting does well then we should expect that players who have negative rank differential will also outperform what we would expect based upon CSS ranking of them and thus have positive metric differential (Δ metric), and vice versa for players with positive rank differential. In general, our approach for the ith draft player can be stated as:

$$\Delta \text{ metric}_i = f_{metric}(\Delta \text{ rank}_i)$$

where $f_{metric}$ is a function to be estimated separately for TOI, GP and GVT.

For TOI, GP and GVT we looked at what each player achieved relative to what we would have expected from them based upon their CSS ranking and compared that to what we would have expected based upon their actual draft selection. To calculate the expected values we used the LOESS regressions from Figure 1. Figure 2 has a plot of each player's first seven season's total TOI minus the Expected TOI for each player plotted against their rank differential as well as the same plot for TOI, GP and GVT. If team scouting was perfect, there would only be players in the upper left and lower right quadrants of these graphs. Players in the upper left are those that exceeded CSS expectations and were taken earlier than CSS had them ranked. Players in the lower left are those that underperformed CSS expectations and teams drafted them earlier than CSS had them ranked. Each graph in Figure 2 also contains a green curve that is the estimated smoothed $f_{metric}$. It is important to note that all of these curves are negatively sloped and roughly pass through the origin (0,0). This suggests that when teams differentiate from the CSS ordering they are, on average, gaining some value in terms of TOI, GP or GVT. To estimate the average net values that team scouting contributes above and beyond the CSS, we estimated the TOI, GP and GVT gained at each rank differential for all players. That is, we calculated:

$$[\Sigma_i \hat{f}_{metric}(\Delta \text{ rank}_i) 1(\Delta \text{ rank}_i > 0) - \Sigma_i \hat{f}_{metric}(\Delta \text{ rank}_i) 1(\Delta \text{ rank}_i < 0)] / [\text{Total number of selections}]$$

which represents the estimated average gained on each metric per draft pick. We then scaled the average values of each to seven selections per season which is the average number of picks that a team has in the current NHL Entry Draft. This quantity then represents per team per draft gain from the information generated by internal scouting. We then used the following information to estimate the average annual revenue differential due to team scouting. Using 20 minutes per game as an average, we estimated that team gained approximately $1.7MM in value from their internal

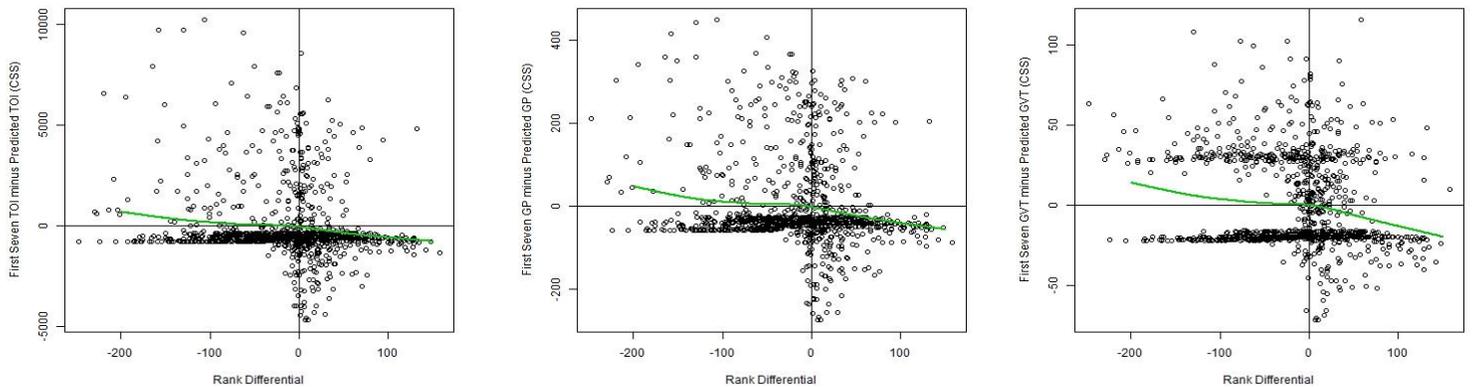

Figure 2: Relationship between Relative Performance and Rank Differential: (l to r) TOI, GP, GVT

scouting. Since the average NHL salary per game is approximately $29,300 (Dorish, 2011), we can say that in terms of GP team scouting adds approximately $2.5MM with each draft. Using GVT, we estimate that team scouting gains a team about $5.3MM in value using the metric that a goal in the NHL is worth approximately $1/3MM (Vollman, 2012). A further analysis by position (including the estimation of $f_{metric}$ by position) yielded similar relationships and results for each of these three metrics.

We also analyzed the value gained over these drafts by team. There are some winners and losers but the average gains per team per draft pick across these five years did not differ significantly from what would have been expected by chance (Shapiro-Wilk p-values: p>0.1 for all). We used an average here since Atlanta, Columbus and Minnesota entered the league between 1999 and 2000 and participated in fewer drafts and, hence, had fewer draft seletions than the other teams. No teams were outlier in this analysis. Further, we looked at the correlation in team average return over CSS for 1998-2000 and the team average return over CSS for 2001-2002. The correlations were slightly positive, r ~0.2 for all three metrics, but not significantly so (p>0.1 for all). No team over this period outperformed the others and no team got a significantly larger or smaller average return from their internal scouting. We further have no way to estimate each team's scouting budget, emphasis on particular leagues or countries, or draft strategies based on positional need or best available talent.

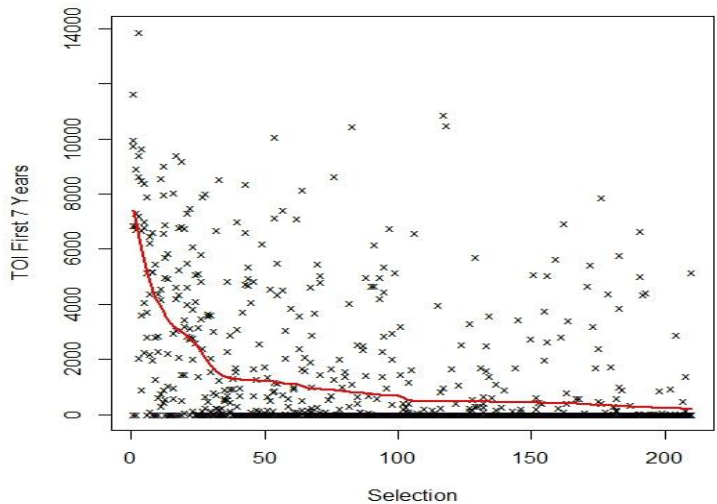

Figure 3: Plot of (Unscaled) Draft Value Pick Chart on Graph of First Seven Years TOI v. Selection

**Conclusions**

In this paper we have looked at the impact of team scouting compared to the NHL's CSS. The metrics for performance that we have used are TOI, GP and GVT for the first seven seasons of a player's career. We began by looking at the chance that the ordering by CSS and by team scouting resulted in choosing optimal or nearly optimal players. At a given selection, teams were not significantly better at picking optimal players with respect to TOI, GP, or GVT but were markedly better at selecting players within a half standard deviation of optimal. There is a clear benefit here to the additional information that team's possess relative to the NHL CSS. To quantify the value of this information we presented a novel methodology that looks at the expected differential from the two rankings relative to the difference in the rankings themselves. This new and comprehensive approach can be applied to *any sport and any method* for ordering draft selections and evaluating the performance of players. It is clear from this analysis that individual NHL team rankings consistently outperform CSS rankings. This should be expected since teams use considerable resources to gain further information about each potential draftee. The models here form the basis for a new NHL Draft Value Pick Chart (Table 4 and Figure 3) based upon the NHL TOI for a given selection in their first seven post-draft seasons. We prefer TOI as it is the most direct of these measures for assessing the value that teams give to their draft choices. For TOI, GP and GVT, we find that the average value that a team gets from their scouting is between $1.8MM and $5.2MM per year. This range represents the average return that teams get on the total amount that they spend on scouting. It also represents a set of bounds on the amounts teams should budget for internal scouting. Further there was variability between teams in their average return from drafting; however, that variability was not beyond what would have been expected by chance. This means that there is not strong enough evidence to suggest that drafting quality is different across teams, at least for the five years that we considered. A larger sample over a longer period of time could provide additional evidence of differences amongst teams.

This analysis could be improved by having better metrics for career player performance. The three that we have used here, TOI, GVT and GP, are reasonable proxies. As hockey analytics develop, utilizing more advanced methods like the Expected Goals Model (EGM), Macdonald (2012) or the Total Hockey Ratings (THoR) (Schuckers and Curro, 2013) will provide better estimates for player value and the value of player scouting once they are available for historical data. In this analysis we focused on five years of NHL Draft Entry selections. Additional years of draft data would provide better estimates of player value and of the value of team scouting. We also note that while team scouting outperforms the Central Scouting Service, team scouting is far from optimal. A future analysis might look at the predictive power of analytics such as league equivalencies, see for example Desjardins (2004) or Vollman (2011), to rank players and compare results from that sort of analysis to those given here. Similarly, better data on teams' front office personnel and duties would provide proxies for team budgets and geographic areas of emphasis. With all that in mind, however, it is clear from this analysis that NHL teams are getting considerable financial benefit from their internal scouting. They are beating the market set by the Central Scouting Service.

APPENDIX

Table 4: NHL DRAFT VALUE PICK CHART

| Selection | Value | Selection | Value | Selection | Value | Selection | Value | Selection | Value | Selection | Value | Selection | Value |
|---|---|---|---|---|---|---|---|---|---|---|---|---|---|
| *1* | 1000 | *31* | 221 | *61* | 150 | *91* | 98 | *121* | 63 | *151* | 58 | *181* | 42 |
| *2* | 932 | *32* | 210 | *62* | 148 | *92* | 96 | *122* | 63 | *152* | 57 | *182* | 42 |
| *3* | 862 | *33* | 200 | *63* | 144 | *93* | 95 | *123* | 63 | *153* | 57 | *183* | 41 |
| *4* | 801 | *34* | 191 | *64* | 140 | *94* | 94 | *124* | 63 | *154* | 57 | *184* | 41 |
| *5* | 743 | *35* | 186 | *65* | 135 | *95* | 93 | *125* | 62 | *155* | 56 | *185* | 40 |
| *6* | 691 | *36* | 180 | *66* | 128 | *96* | 92 | *126* | 62 | *156* | 56 | *186* | 39 |
| *7* | 645 | *37* | 176 | *67* | 127 | *97* | 92 | *127* | 62 | *157* | 56 | *187* | 38 |
| *8* | 606 | *38* | 172 | *68* | 126 | *98* | 91 | *128* | 62 | *158* | 56 | *188* | 38 |
| *9* | 567 | *39* | 171 | *69* | 126 | *99* | 91 | *129* | 62 | *159* | 55 | *189* | 37 |
| *10* | 546 | *40* | 171 | *70* | 125 | *100* | 91 | *130* | 62 | *160* | 55 | *190* | 36 |
| *11* | 523 | *41* | 170 | *71* | 124 | *101* | 89 | *131* | 62 | *161* | 54 | *191* | 35 |
| *12* | 502 | *42* | 169 | *72* | 123 | *102* | 81 | *132* | 62 | *162* | 54 | *192* | 35 |
| *13* | 482 | *43* | 168 | *73* | 122 | *103* | 73 | *133* | 62 | *163* | 54 | *193* | 34 |
| *14* | 462 | *44* | 168 | *74* | 121 | *104* | 67 | *134* | 62 | *164* | 53 | *194* | 34 |
| *15* | 446 | *45* | 167 | *75* | 119 | *105* | 67 | *135* | 62 | *165* | 53 | *195* | 34 |
| *16* | 431 | *46* | 166 | *76* | 118 | *106* | 66 | *136* | 61 | *166* | 53 | *196* | 33 |
| *17* | 420 | *47* | 165 | *77* | 117 | *107* | 66 | *137* | 61 | *167* | 53 | *197* | 33 |
| *18* | 410 | *48* | 165 | *78* | 116 | *108* | 65 | *138* | 61 | *168* | 52 | *198* | 33 |
| *19* | 401 | *49* | 164 | *79* | 115 | *109* | 65 | *139* | 61 | *169* | 51 | *199* | 32 |
| *20* | 391 | *50* | 163 | *80* | 113 | *110* | 65 | *140* | 61 | *170* | 50 | *200* | 32 |
| *21* | 381 | *51* | 162 | *81* | 111 | *111* | 65 | *141* | 61 | *171* | 49 | *201* | 31 |
| *22* | 371 | *52* | 162 | *82* | 110 | *112* | 65 | *142* | 61 | *172* | 48 | *202* | 31 |
| *23* | 357 | *53* | 161 | *83* | 109 | *113* | 65 | *143* | 61 | *173* | 47 | *203* | 30 |
| *24* | 341 | *54* | 159 | *84* | 107 | *114* | 65 | *144* | 60 | *174* | 46 | *204* | 30 |
| *25* | 325 | *55* | 155 | *85* | 107 | *115* | 65 | *145* | 60 | *175* | 45 | *205* | 30 |
| *26* | 311 | *56* | 152 | *86* | 105 | *116* | 65 | *146* | 60 | *176* | 45 | *206* | 29 |
| *27* | 296 | *57* | 151 | *87* | 104 | *117* | 65 | *147* | 59 | *177* | 44 | *207* | 28 |
| *28* | 276 | *58* | 151 | *88* | 103 | *118* | 64 | *148* | 59 | *178* | 44 | *208* | 27 |
| *29* | 257 | *59* | 150 | *89* | 101 | *119* | 64 | *149* | 58 | *179* | 43 | *209* | 26 |
| *30* | 238 | *60* | 150 | *90* | 100 | *120* | 64 | *150* | 58 | *180* | 43 | *210* | 25 |

Note that we used a monotonic regression of TOI onto Selection for this chart following the `monreg` package in R.